\documentclass{aa}

\usepackage[T1]{fontenc}
\usepackage{pslatex}
\usepackage{amsmath}
\usepackage{amsfonts}
\usepackage{xcolor}
\usepackage{url}
\usepackage{bm}
\usepackage{graphicx}
\usepackage{amssymb}
\usepackage{soul}
\usepackage{booktabs}
\usepackage{array}
\usepackage[utf8]{inputenc}
\inputencoding{latin1}
\inputencoding{utf8}
\usepackage{appendix}

\def\lsim{ \lower .75ex\hbox{$\sim$} \llap{\raise .27ex \hbox{$<$}} }
\def\gsim{ \lower .75ex \hbox{$\sim$} \llap{\raise .27ex \hbox{$>$}} }

\newcommand{\fig}[1]{Fig.~\ref{fig:#1}}

\newcommand{\bi}{\begin{itemize}}
\newcommand{\ei}{\end{itemize}}

\DeclareMathOperator{\sech}{sech}
\usepackage[colorlinks=true,linkcolor=blue,citecolor=blue]{hyperref}

\title{FR0 jets and recollimation-induced instabilities} 

\author{
A. Costa\inst{1,2}
\and G. Bodo\inst{3}
\and F. Tavecchio\inst{2}
\and P. Rossi\inst{3}
\and A. Capetti\inst{3}
\and S. Massaglia\inst{4}
\and A. Sciaccaluga\inst{5,2}
\and R.~D. Baldi\inst{6}
\and G.~Giovannini\inst{6,7}
}

\institute{
DiSAT, Università dell’Insubria, Via Valleggio 11, I-22100 Como, Italy\\
\email{agnese.costa@inaf.it}
\and
INAF -- Osservatorio Astronomico di Brera, via E. Bianchi 46, 23807 Merate, Italy
\and
INAF -- Osservatorio Astrofisico di Torino, Strada Osservatorio 20, 10025 Pino Torinese, Italy
\and
Dipartimento di Fisica, Università degli Studi di Torino, Via Pietro Giuria 1, 10125 Torino, Italy
\and
Dipartimento di Fisica, Universita` degli Studi di Genova, Via Dodecaneso 33, I-16146 Genova, Italy
\and
INAF - Istituto di Radioastronomia, Via P. Gobetti 101, I-40129 Bologna, Italy
\and
Dipartimento di Fisica e Astronomia, Università di Bologna, Via P. Gobetti 93, I-40129 Bologna, Italy
}
\date{}

\begin{document}



\abstract
{The recently discovered population of faint FR0 radiogalaxies has been interpreted as the extension to low power of the classical FRI sources. Their radio emission appears to be concentrated in very compact (pc-scale) cores, any extended emission is very weak or absent and VLBI observations show that jets are already mildly or sub-relativistic at pc scales. Based on these  observational properties we propose here that the jets of FR0s are strongly decelerated and disturbed at pc scale by hydrodynamical instabilities.}{With the above scenario in mind, we study the dynamics of a low-power relativistic jet propagating into a confining external medium, focusing on the effects of entrainment and mixing promoted by the instabilities developing at the jet-environment interface downstream of a recollimation shock.}{We perform a 3D relativistic hydrodynamical simulation of a recollimated jet by means of the state-of-the-art code PLUTO. The jet is initially conical, relativistic (with initial Lorentz Factor $\Gamma$=5), cold and light with respect to the confining medium, whose pressure is assumed to slowly decline with distance. The magnetic field is assumed to be dynamically unimportant.}{The 3D simulation shows that, after the first recollimation/reflection shock system, a rapidly growing instability develops, as a result of
the interplay between recollimation-induced instabilities and Richtmyer-Meshkov modes.
In turn, the instabilities promote strong mixing and entrainment that rapidly lead to the deceleration of the jet and spread its momentum to slowly moving, highly turbulent external gas. We argue that this mechanism could account for the peculiarities of the low-power FR0 jets. For outflows with higher power, Lorentz factor or magnetic field, we expect that the destabilizing effects are less effective, allowing the survival of the jet up to the kpc scale, as observed in FRIs.}{}{}


\maketitle

\section{Introduction}
Despite decades of efforts, the comprehension of relativistic jets ejected by active galactic nuclei (AGN) is still rather sketchy \citep[e.g.][]{Blandford19}. Challenges for current research concern the mechanisms able to accelerate and collimate the jet in the central supermassive black hole (SMBH) vicinity, the composition of the outflowing plasma (pair {\it vs} proton-electron, matter {\it vs} magnetically dominated), the velocity structure of the flow, the mechanism(s) accelerating particles to ultra-relativistic energies. Since their discovery, one of the most active research topics has been the role of the various instabilities in shaping the jet's dynamical and dissipative properties. A very broad range of instabilities  can be at work: Kelvin-Helmholtz, current driven, pressure driven, centrifugal, Rayleigh-Taylor \citep[e.g.][]{birkinshaw96, Bodo13, Bodo19, Kim17, Kim18, Begelman98, Begelman19, gourgouliatos18}, and these different instabilities may lead to different outcomes in the jet dynamics and energy dissipation. In particular, they are thought to play a prominent role in low-power jets (classified as Fanaroff-Riley I (FRI) sources, \citealt{FR74}), whose structure has been generally interpreted as the result of instabilities and subsequent entrainment of external gas \citep[e.g.][]{rossi08, laing14, perucho14,massaglia16, rossi20}, eventually disrupting the jet at kpc scales. On the other hand, powerful FRII jets seem to be much less prone to instabilities and can reach distances up to Mpc scale  \citep{willis74}, where they feed the giant radio lobes.

It has been recently realized that, among local extragalactic jetted sources, the largest fraction is composed by low-power objects, with radio morphology characterized by a compact core with virtually no extended emission at kpc scale (see \citealt{baldi23}, and references therein). Since the radio properties make this population the natural low-power extension of the FRI classical radio galaxies \citep{FR74}, they have been dubbed ``FR0s" \citep{ghisellini11,sadler14}. VLBI studies of FR0s \citep{cheng18,cheng21,baldi21b,giovannini23} reported that the complex jet structure and the large number of two-sided structures are strong evidence that, contrary to what is observed in FRI radio galaxies, FR0 jets in VLBI images are mildly or sub-relativistic, with bulk velocity on the order of $0.5c$ or less at parsec scales. Moreover many FR0 jets are complex and display substructures on pc scales, hinting for a strong interaction with the surrounding interstellar medium. In turn, this suggests that low-power jets of FR0s are possibly able to efficiently remove cold gas from the nucleus of the host galaxy, thus influencing the accretion onto the SMBH \citep{baldi23}. In fact, observational evidence continues to mount that low-power jetted AGNs in general can deposit a large amount of jet energy in the interstellar medium through shocks and turbulence (e.g., \citealt{venturi21,pareira22,nandi23}). The lack of a relevant extended emission and the morphology observed at VLBI scales indicate that the deceleration and the dissipation of the FR0 jet power occur close to the SMBH, at pc scale, therefore challenging any scenario involving mechanisms operating in a smooth and gradual way. Furthermore, the compact core implies the existence of a localized region of intense dissipation close to the AGN core. 

Building on this observational base, we adopt here a scenario for the dynamics of FR0 jets whose main actor is a recollimation shock. 
Specifically, we assume that a (weakly magnetized, low-power) jet expands and rapidly becomes underpressured with respect to the external gas. In such situation a recollimation shock structure develops into the jet \citep[e.g.][]{KomissarovFalle97,BodoTavecchio18}, accompanied by growing instabilities that decelerate and perturb the jet and quickly destroy the flow. In this framework the shock plays the double role of ensuring both the dissipation (and the subsequent emission) of part of the jet kinetic power, and the excitation of instabilities which rapidly decelerate and disrupt the jet.

Recent simulations show that recollimation shocks promote the formation of instabilities at the jet/external medium interface, that can have a strong impact on the flow. Several studies, in particular, concentrate on the Rayleigh-Taylor (RTI, \citealt{Matsumoto13,Matsumoto19,GottliebHD}), centrifugal (CFI, induced by the effective gravity resulting from the motion of the plasma along curved streamlines in the recollimation region \citealt{gourgouliatos18}) and Richtmyer-Meshkov (RMI; triggered by the passage of the reflected shock at the jet/external medium interface, e.g. \citealt{Matsumoto13,Matsumoto19,GottliebHD}) instabilities. Here we will focus on the case of a ``light" jet, i.e. stable against the RTI \citep[e.g.][]{abolmasov23}.
Instabilities excited by recollimation can be damped by a sufficiently intense magnetic field with a suitable geometry \citep[]{Matsumoto21,GottliebMHD}, but in this preliminary exploration we assume a pure hydrodynamical (HD) jet, a choice suitable to model a flow in which the magnetic field is not dynamically important.

\section{Simulation}

As mentioned in the Introduction, we consider a scenario in which the jet, after an initial phase of free expansion, becomes underpressured with respect to the ambient medium. In such situation, if we assume axisymmetry, the jet is characterized by a series of recollimation and reflection shocks \citep{KomissarovFalle97}, as confirmed by two-dimensional cylindrical simulations \citep[e.g.][]{Mizuno15}. Our aim is to investigate the stability of such configuration, when the axisymmetry constraint is relaxed, in agreement with the real jet images \citep[see][and references therein]{boccardi21}{}{}. To this aim we first perform two-dimensional simulations starting with a conical jet (as in \citealt{BodoTavecchio18}) and let the system evolve until a steady state is reached. We then use this steady state as initial condition for the 3D simulations.

The jet, whose initial opening angle is $\theta_j = 0.2$, is relativistic, with a Lorentz factor $\Gamma_j = 5$ at injection, and propagates through a surrounding isothermal medium at rest, with density and pressure that decay along $z$ with power law profiles of index $\eta =0.5$. The jet is injected at a distance $z_0$ from the cone vertex. We simulate a "light'' jet, which is under-dense and under-pressured with respect to the confining gas. The values of the density and pressure ratios between jet and ambient, at the jet base, are respectively:
\begin{equation}
 \frac{\rho_{j,0}}{\rho_{ext,0}} = 7.6 \times 10^{-6} \,,
 \qquad
 \frac{p_{j,0}}{p_{ext,0}} = 10^{-3}\,,
 \label{eq:jet}
\end{equation}
and the external pressure is:
\begin{equation}
\frac{p_{ext,0}}{\rho_{ext,0} \, c^2}=3 \times 10^{-6},
\label{eq:extT}
\end{equation}
corresponding to a temperature of $3 \times 10^7$K.

We perform the simulations with the relativistic hydrodynamical (RHD) module of the state-of-the-art code PLUTO \citep[][]{MignonePLUTO}. The computational box covers the domain $[-5,5]\times[-5,5]\times[1,30]$ in units of $z_0$ and we adopt a resolution of 35 points per initial jet radius $r_0 = 0.2 z_0$. We set outflow conditions at all boundaries except at $z=z_0$, where we fix the injection of the jet for $r < r_0$ and the environment static profiles. We run the simulation up to $t_f = 368\, z_0/c$ (corresponding to $1840\, r_0/c$), when a quasi steady-state is reached. We complement the RHD equations with the equation for the evolution of a passive tracer, that is set unity for the injected jet material and to zero for the ambient medium. In this way we can study in detail the mixing process between jet and ambient. In our simulations we adopt units so that $c=1$, $z_0=1$ and $\rho_{0,ext}=1$. The unit time will be $t_0 = z_0/c$. See the Appendix for more details on the numerical setup.

\section{Results}
\begin{figure}
\begin{center}
\includegraphics[scale=0.4]{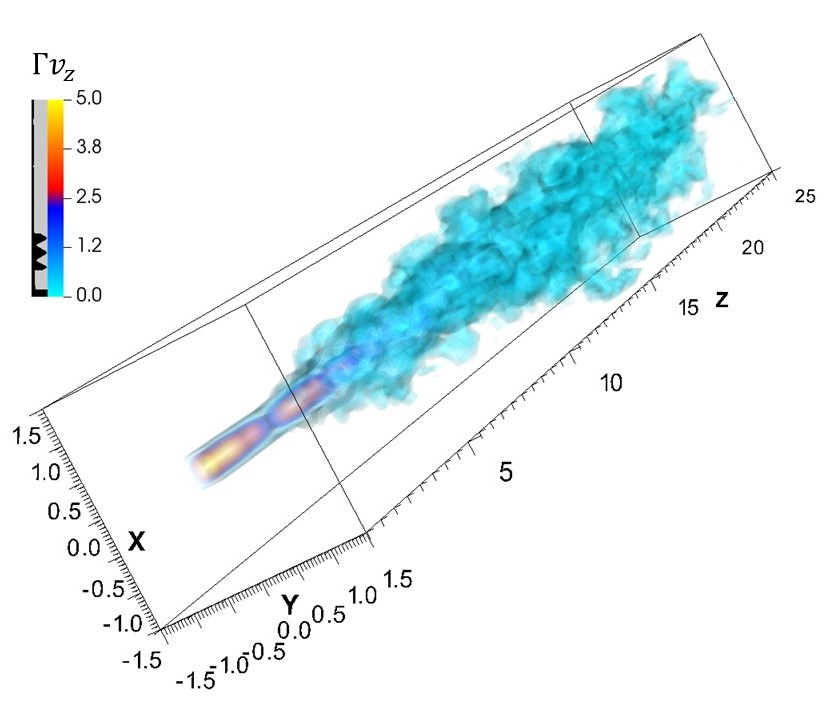}
\caption{3D volume rendering of the $z$ component of the jet  4-velocity . The black-grey bar next to the colorbar shows which values of $\Gamma v_z$ are opaque (grey), and which are transparent (black).}
\label{fig:3D}
\end{center}
\end{figure}

In the 3D simulations, relaxing the axisymmetry constraint, the dynamics is strongly modified by instabilities that show a very fast growth. The global structure of the perturbed jet is presented in figure \ref{fig:3D}, where we display a 3D volume rendering  of the $z$ component of the 4-velocity $\Gamma v_z$ at $t_f$. At the base, the jet is relativistic ($\Gamma v_z >3$, yellow-red), and it is possible to distinguish the first compression stage, followed by an expansion and a second compression phase. The jet later decelerates while it entrains external material ($\Gamma v_z \leq 1$, light blue), becoming quickly sub-relativistic. 
The jet instability seems to be ascribable to a combination of the Kelvin-Helmholtz instability (KHI), the CFI, and the RMI instabilities. In the slow, entrainment regions of Fig. \ref{fig:3D}, it is possible to find signatures of KHI-induced helical deformation, but it is the RMI that develops important non-linear perturbations.

\begin{figure}
\centering
\includegraphics[scale=0.535]{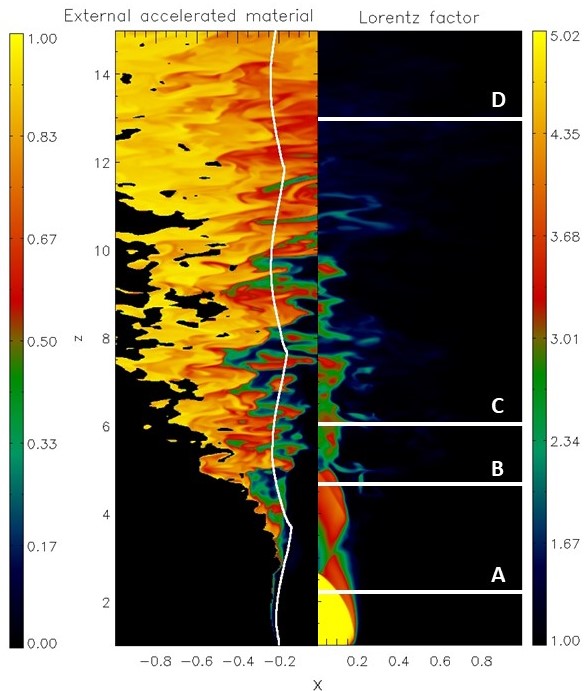}
\caption{The two pictures display the maps of the Lorentz factor (right panel) and of the tracer of the medium material (left panel) that is accelerated above a threshold of $v_z \geq 0.1$ in the $x-z$ plane, at $y=0$. The white curve on the left is a contour of the 2D stationary jet. White horizontal lines on the right indicate 4 different distances $z*$ at which we evaluate the $x-y$ cuts in Fig. \ref{fig:tagli_xy}.}
\label{fig:xz}
\end{figure}

\begin{figure}
\centering
\includegraphics[scale=0.4]{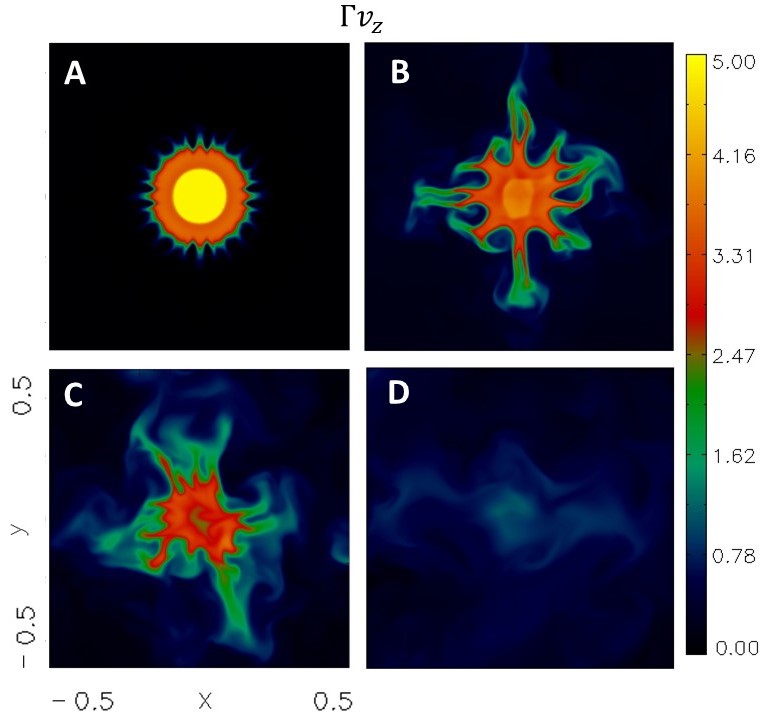}
\caption{The 4-velocity $\Gamma v_z$ in the $x-y$ plane at different $z*=2.3,\,4.7,\,6.1,\,13 $ respectively for A, B, C, and D defined inn Fig.~\ref{fig:xz}.}
\label{fig:tagli_xy}
\end{figure}

More details on the dynamics can be obtained from Figures \ref{fig:xz} and \ref{fig:tagli_xy}. In Fig. \ref{fig:xz} we show two-dimensional slices, at $y=0$, of the distribution of the Lorentz factor (right) and of the tracer of the external material, moving with $v_z>0.1 $ (left). The tracer value represents the fraction of external material present in a given computational cell. The stationary axisymmetric jet profile is over-plotted on the left in white with a contour of the jet tracer. The four white horizontal lines on the right panel indicate the locations of the $z=z^*$ cuts shown in Fig. \ref{fig:tagli_xy}, where we represent the $z$ component of the four-velocity $\Gamma v_z$. 

The jet is decelerated by the first strong recollimation shock (located at the boundary of the yellow region in \fig{xz}), that reaches the axis at $z\simeq 2.6$ and is reflected, reaching  an anti-node at $z\simeq 3.8$. During the expansion stage, after $z\simeq 3$, some external material starts to be entrained at the jet-environment contact discontinuity (CD) (see \fig{xz}), because of small scale perturbations induced by the recollimation instabilities. A secondary effect of these perturbations is that the reflection shock becomes able to cross the CD where it is corrugated, and excites the RMI \citep[][]{Matsumoto13,Matsumoto19}. The jet finally becomes unstable to RMI approximately at the anti-node, as it can be seen in \fig{tagli_xy}, panel B, a region particularly favourable to the starting of instabilities \citep[][]{hydro}.
After the anti-node, the jet is recollimated through a second shock, but the downstream region is unstable, and after the second recollimation point, at $z\simeq 5$, the jet is not able to expand again. Only a portion of the material in the jet core continues to propagate at a relativistic velocity, with $\Gamma \simeq 3$ up to $z\simeq10$, while entraining external medium through the turbulent interaction between the two fluids. Finally, at $z>10$, the jet becomes sub-relativistic. These different stages are also displayed in Fig. \ref{fig:tagli_xy}: panel A shows the transversal jet structure before the first recollimation point, where we clearly distinguish the unshocked jet portion in yellow and the shocked portion in orange; panels B and C show the development of perturbations and the progressive deceleration of the jet core surrounded by a slow mixing layer; in panel D we see that all the jet has become sub-relativistic.

\begin{figure}
\centering
\includegraphics[scale=0.5]{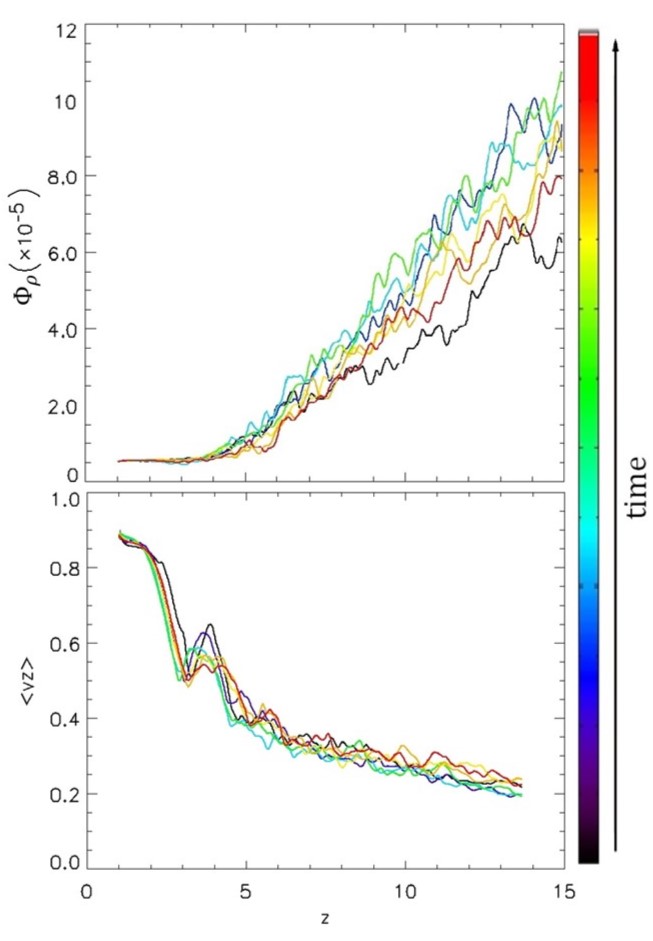}
\caption{Upper panel: mass flux, $\Phi_\rho(z,t)=\int_{xy}\rho \Gamma v_z dx dy$, starting from $\Phi_\rho (z_0) = 5.3\times 10^{-6}$, as function of $z$ at different late times $t=[140,180,220,260,300,340,368]$. Lower panel: averaged propagation velocity. More details can be found in the appendix.}
\label{fig:entrainment}
\end{figure}
\begin{figure}
\centering
\includegraphics[scale=0.5]{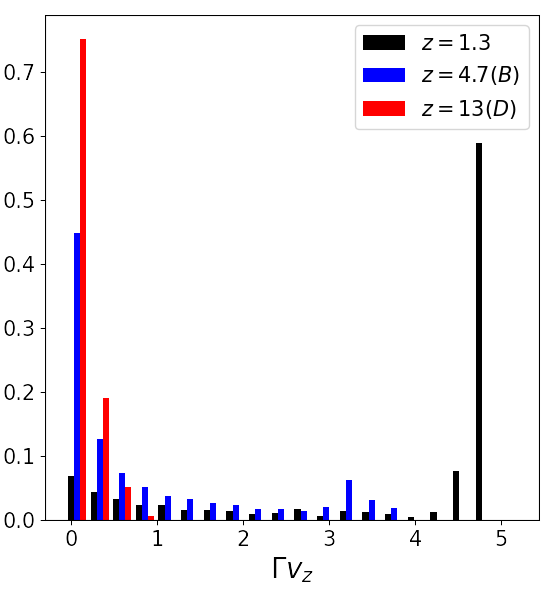}
\caption{The histogram displays the fraction of cells with a given jet velocity $\Gamma v_z$ at the end of the simulation, calculated on three $z=z*$ planes, representing the initial jet, the jet after the anti-node (as in Fig. \ref{fig:tagli_xy}B) that has entrained material, and the jet further in $z$, where it has become sub-relativistic (as in Fig. \ref{fig:tagli_xy}D).}
\label{fig:histopy}
\end{figure}

\begin{figure}
\centering
\includegraphics[scale=0.29]{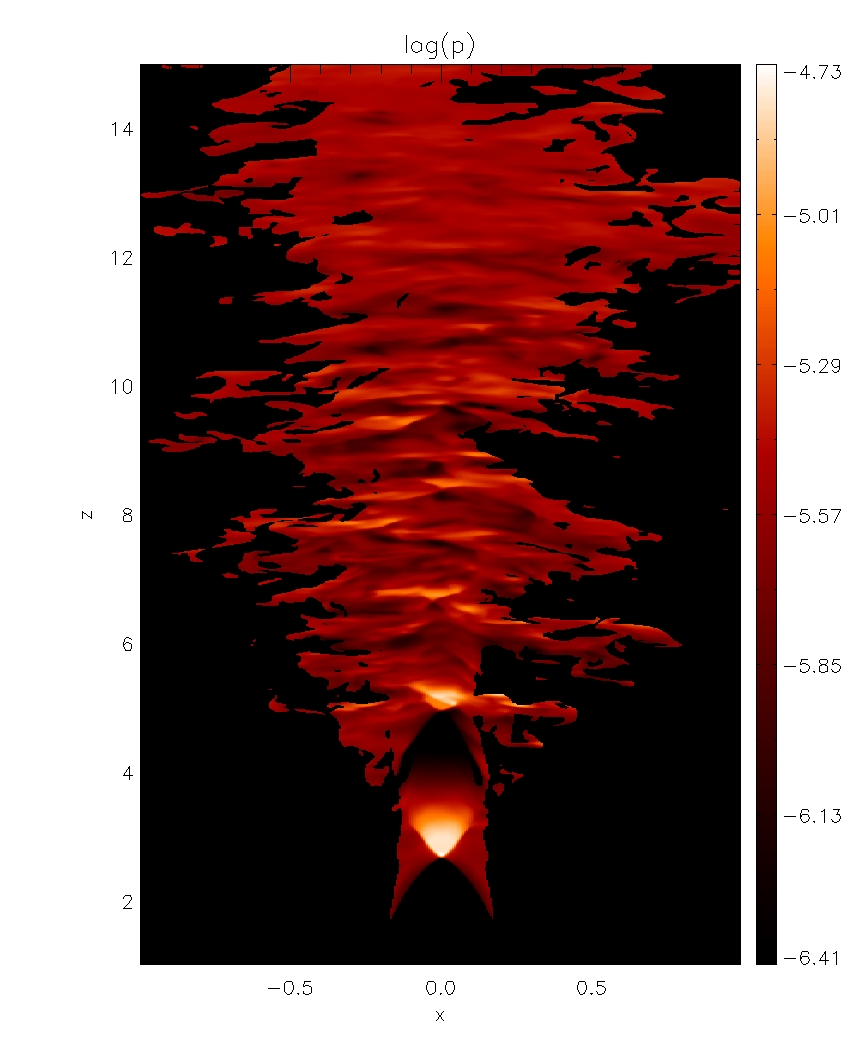}
\caption{The picture shows the $y=0$ slice of the logarithm of the pressure in code units, $\mbox{log}(p)$. The colder regions, with $\mathcal{T} < 0.1$, are not displayed. }
\label{fig:prs}
\end{figure}
Figure \ref{fig:entrainment} compares the entrainment of external material with the overall jet deceleration, by plotting the mass flux $\Phi_\rho(z,t)$ and the average velocity $\langle v_z(z,t)\rangle$ in the $z$ direction, as functions of the altitude $z$, where the different lines refer to different times, from black to red, at $t_f$. The top panel shows $\Phi_\rho(z) = \int_{xy}\rho\Gamma v_z\,dxdy$. For small values of $z$ the flux is small, since the jet density is low ($\rho=7.6 \times 10^{-6}$), but it starts increasing at $z\simeq 3$, where there are the first signatures of entrainment. After the second recollimation point, at $z\simeq 5$, the flux grows significantly, as a result of the entrainment of the heavy external medium. The increase in the flux is almost linear (\fig{xz}).
After increasing with time at the beginning of the simulation, as shown from curves from $t=140$ to $t=260$, the mass flux seems to converge to a stationary profile, showing little dispersion from $t=300$ to $t_f$.

The bottom panel of Fig.~\ref{fig:entrainment} displays the propagation velocity averaged on $x-y$ planes.
The strong decrease of $\langle v_z \rangle$ up to $z\simeq 3$ is due to the recollimation shock (see also the red region in Fig. \ref{fig:tagli_xy}, panel A); then the jet accelerates again as it expands. After the anti-node at $z\simeq3.8$, the average velocity decreases as a result of both the second recollimation shock and of the starting of the entrainment process. 
After $z\simeq 5$, the jet keeps slowing down, because part of its momentum is transferred to the heavier entrained material, reaching sub-relativistic velocities in a smooth way (see also Fig. \ref{fig:tagli_xy}, panel D). The curves for different times almost overlap, with variations $< 10\%$, indicating that the jet has reached an almost quasi steady-state.
The transition from relativistic to sub-relativistic velocities is also displayed in Fig. \ref{fig:histopy}, showing an histogram of $\Gamma v_z$, the $z$ component of the four-velocity distribution, in the $x-y$ plane, at three different positions in $z$. 
Black refers to the jet base, blue refers to the region just after the second recollimation point (see also Fig. \ref{fig:tagli_xy}, panel B), where there still is a fast spine, and red to the region where the flow is completely sub-relativistic (see also Fig. \ref{fig:tagli_xy}, panel D).

\fig{prs} shows a pressure map (in  units of $\rho_{ext,0}c^2$, in logarithmic scale) in the $x-z$ plane at $y=0$, at the end of the simulation. The cold regions for which $\mathcal{T}= p/\rho c^2 <0.1$, like the injection and the environment, are left black in the image.
Initially the jet is cold and its energy is mainly kinetic but, when the fluid crosses the shocks, its thermal energy increases, resulting in large values of temperature in the shock downstream. After the first recollimation shock, the jet reaches  pressure balance with the cold environment ($\mathcal{T}_{ext,0}=3\times 10^{-6}$, from Eq. \ref{eq:extT}); pressure is then further increased downstream of the reflection shock, for $z>2.6$. After the expansion, we observe another highly pressurized region after the second recollimation point at $z\simeq5$. In these regions we expect the presence of non-thermal particles accelerated at the shocks and emitting synchrotron radiation. The synchrotron emissivity can be qualitatively estimated as $j\propto pB^2$ \citep[e.g.][]{BodoTavecchio18}, where $p$ and $B$ are the pressure and the magnetic field strength in the jet. Our simulation is purely hydrodynamical, hence we are not able to infer the magnetic field strength.
Assuming that the magnetic field energy density is a fraction of the thermal pressure, the emissivity will scale as $j\propto p^2$. We therefore expect that the regions downstream of the reflection shocks are the brightest components, with an emissivity larger by a factor of $10-100$ with respect to the other regions, which may originate much weaker diffuse emission.

\section{Discussion}

We have reported the results of a simulation of a conical jet, underpressured with respect to the environment, in which the development of an oblique recollimation shock promotes the growth of instabilities. The instability evolution leads to vigorous turbulent mixing with the external gas which is entrained by the jet, and the consequent spreading of momentum results in a rapid jet deceleration. Hydrodynamical simulations are scale invariant in principle; here we set the length and density scales, that represent the jet distance from the engine  ($z_0$) and the external density ($\rho_{ext,0}$), 
to values inferred from the typical properties of low-power radio galaxies \citep[e.g.][]{heckman14, russell15, boccardi21, casadio21}{}{}. Assuming $z_0=1$ pc and $\rho_{ext,0}=1 \, m_p\mbox{cm}^{-3}$ the simulation is suitable to reproduce the observed properties of FR0 sources \citep[e.g.][]{baldi23}{}{}. The simulated jet is confined at a distance of $\sim 3$ pc, where it crosses powerful standing shocks that are sites of particle acceleration and non-thermal emission, 
creating bright spots that will result in the observed core at pc scales \citep[][]{baldi21b}{}{}. The jet is mildly relativistic up to a distance of the order of $10$ pc, transferring its momentum to entrained external material, to finally become sub-relativistic on larger scales. 
This is in agreement with the observed properties of FR0s. 
\citep[e.g.][]{baldi15,baldi19,cheng18,capetti20a,cheng21,giovannini23}{}{}.
The injected jet luminosity is:
\begin{align}
    L_j = & \pi (z_0 \theta_j)^2 v_{z,j} \rho_{j,0} c^2h_{j,0} \Gamma^2 \\
    \simeq & 10^{40}\left(\frac{\rho_{ext,0}}{1\,m_p\,\mbox{cm}^{-3}}\right)\left(\frac{z_0}{1\,\mbox{pc}}\right)^2\,\mbox{erg s$^{-1}$},
\end{align}
where $h$ is the specific enthalpy and in the initially cold jet $h_{j,0}\simeq 1$. This jet power is consistent with typical values derived for FR0s \citep[][]{baldi18}{}{}. Nevertheless, this simulation should be interpreted as a case study of a light and cold jet. We expect lighter setups, with $L_j\leq10^{40}\,\mbox{erg s}^{-1}$ (thus describing low-power FR0 jets), to be even more unstable.

It is well known that the accretion flow and the environment play a key role in determining the confinement properties of jets, and their transition to a conical and cylindrical shape \citep[e.g.][]{permar07, park23, rohoza23}{}{}. In particular, FR0s are hosted by giant elliptical galaxies, with hints of a hot corona \citep[][]{hardcastle07, torresi18}{}{} from X-ray spectroscopic data. Observations of low-power radio galaxies suggest that jets propagate with a conical geometry at scales of $1-20$ pc in a stratified medium \citep[e.g.][]{russell15, boccardi21, casadio21}{}{}, with evidences of recollimation shocks \citep[for BL Lac][]{casadio21}{}{}.   

In terms of luminosity, FR0s can be interpreted as the low-power tail of the FRI class of radio-galaxies, with $L\leq 10^{40} \,\mbox{erg s}^{-1}$ \citep[][]{baldi18}{}{}. In the scenario proposed here, the initial opening angle of the jet, the jet-environment density and pressure ratios play the most important roles in causing the confinement and triggering the instabilities downstream of the shocks. We expect that FRIs, and especially FRIIs, are characterized by more powerful jets, with higher density and/or magnetization, ensuring the stability required to survive the recollimation instabilities and reach the kpc scale. This is consistent with the view proposed in \cite{gourgouliatos18}. The discussion of the role of the different parameters will be addressed in a forthcoming paper (Costa et al. in prep.). More fundamentally, these parameters could be connected to some of the key properties of the central engine, some not directly observable at present, such as the SMBH/accreting disk spin \citep[][]{garosin19,grandi21,giovannini23,lalakos23}{}{}.

\section*{Acknowledgments}
We thank P. Coppi for useful discussions. We aknowledge financial support by a INAF Theory Grant 2022 (PI F. Tavecchio) and the PRIN 2022 (2022C9TNNX) project. We acknowledge support by CINECA, through ISCRA and Accordo Quadro INAF-CINECA, and by PLEIADI, INAF - USC VIII, for the availability of HPC resources. This work has been funded by the EU - Next Generation EU.

\bibliographystyle{aa}
\bibliography{tavecchio}

\appendix
\section{Numerical setup}
The simulations presented in this letter are performed with the PLUTO \citep[][]{MignonePLUTO}{}{} modular code that solves the set of conservation equations of fluid dynamics. In particular we employed the RHD module, that solves the system of relativistic hydrodynamics equations

\begin{equation}
    \frac{\partial}{\partial t}\left(  \begin{matrix}
\rho \Gamma\\
\rho h \Gamma^2 \mathbf{v} \\
\rho h \Gamma^2-p  \\
\rho \Gamma f
\end{matrix}\right) + \mathbf{\nabla}\cdot\left( \begin{matrix}
\rho \Gamma \mathbf{v}\\
\rho h \Gamma^2 \mathbf{v}\mathbf{v}+p\mathbf{I} \\
\rho h \Gamma^2\mathbf{v}  \\
\rho \Gamma f\mathbf{v}
\end{matrix} \right) = \left( \begin{matrix}
0\\
\mathbf{f}_g \\
\mathbf{f}_g\cdot \mathbf{v}  \\
0
\end{matrix}  \right),
\end{equation}
where $\rho,\,\Gamma,\,h,\,\mathbf{v},\,p,$ respectively represent the rest-frame number density, the Lorentz factor, the proper specific enthalpy, the 3-velocity in units of $c$, and the pressure of the fluid. The acceleration term $\mathbf{f}_g$ is the  specific external force three-vector, set to  $\mathbf{f}_g = \nabla p_{ext}(t=0)$, in order to maintain in dynamical equilibrium the static ambient medium.
We also evolve a passive tracer $f$, initially set to $0$ for the surrounding medium, and to $1$ for the injected relativistic jet, to track the evolution of the jet material and to study the mixing between the two fluids.
We close the set of equations with the Taub-Matthews equation of state: 
\begin{equation}
    h = \frac{5}{2}\mathcal{T}+\sqrt{\frac{9}{4}\mathcal{T}^2+1},
\end{equation}
with $\mathcal{T}=p/(\rho c^2)$, that approximates the Synge EoS of a single-species relativistic perfect fluid \citep[][]{MignoneTM}{}{}.
We adopt a linear reconstruction scheme with a second order Runge-Kutta method for time integration, and the HLLC Riemann solver \citep[][]{MignoneHLLC}. 

\subsection{2D setup}
The preliminary axisymmetric simulation is run in 2D cylindrical coordinates $(r,z)$ in a domain $[0,20]\times[0.5,30]$, where the lengths are in units of $z_0$, that represents the distance from the jet launching site. The grid is uniform with $1000\times3000$ points in $[0,1.5]\times[0.5,20]$ and geometrically stretched with $400\times 700$ grid points in the outer regions. 

As initial condition, at $t=0$, in the region $r/z < 0.2$, we have an axisymmetric conical outflow, with opening angle $\theta_j = 0.2$, with Lorentz factor $\Gamma_j=5$, more precisely the velocity components are defined as
\begin{equation}\label{eq:vz}
  v_z = \sqrt{1- \frac{1}{\Gamma_j^2}} \frac{z}{R},   
\end{equation}
\begin{equation}\label{eq:vr}
v_r = \sqrt{1- \frac{1}{\Gamma_j^2}} \frac{r}{R},    
\end{equation}
 where $R$ is the spherical radius defined as $R=\sqrt{r^2+z^2}$.  Density and pressure decay adiabatically  as
\begin{equation}\label{eq:rho}
  \rho_j(r,z,t=0) = \rho_{j}(0,z_0,0) \left(\frac{R}{R_0}\right)^{-2},  
\end{equation}
and pressure
\begin{equation}\label{eq:prs}
 p_j(r,z,t=0) = p_{j}(0,z_0,0) \left(\frac{R}{R_0}\right)^{-2\gamma},
\end{equation}
where $\gamma$ is the adiabatic index derived from the EoS, which in the case of a cold gas yields the classical $\gamma=5/3$.
For $r/z > 0.2$ there is a static medium, whose density $\rho_{ext}$ and pressure are power law functions of the altitude $z$:
\begin{align}
  \rho_{ext}(z,t=0)=\rho_{ext}(z_0,0)\left(\frac{z}{z_0}\right)^{-\eta}, \\
   p_{ext}(z,t=0) = p_{ext}(z_0,0) \left(\frac{z}{z_0}\right)^{-\eta},
\end{align}
where the power law index is $\eta=0.5$. The exact value of $\eta$ is not decisive, as long as it's smaller than $2$, so that the higher pressure of the external medium confines the relativistic jet.
 
We simulate a cold jet, that is light compared to the environment, so we define its density and pressure with respect to the external values, as 
\begin{equation}
    \frac{\rho_{j}(0,z_0,0)}{\rho_{ext}(z_0,0)} = 7.6 \times 10^{-6},\\
     \frac{p_{j}(0,z_0,0)}{p_{ext}(z_0,0)} = 10^{-3},
\end{equation} 
while the external pressure is defined through the temperature: $\mathcal{T}_{ext}(z_0,0) = 3 \times 10^{-6}$. 
In our simulations we set $\rho_{ext,0}$ as the density unit.

In order to avoid numerical noise at the contact discontinuity, the initial condition is smoothed at the jet-environment boundary. We smoothed the Lorentz factor, the density and the pressure, with functions of the type  
\begin{equation}
    q = q_{ext} + \left(q_j-q_{ext}\right)\sech\left[\left(\frac{r}{z\theta_q}\right)^{\alpha_q}\right]     
\end{equation}
in the inlet regions \citep[][]{mukherjee20}{}{}. The index $\alpha_q$ and the angle $\theta_q$, that define the width and the radial scale of the smoothing, depend on the specific quantity $q$, and are different for different values to avoid artificial local extrema in the energy and/or momentum \citep[][]{abolmasov23}{}{}. We set $\theta_\Gamma=0.16$ and $\alpha_\Gamma=8$ for the Lorentz factor, and $\theta_{\rho,p}=0.29$, $\alpha_{\rho,p}=10$ for the density and the pressure.

We use outflow conditions at the right ($r=20$) and upper ($z=30$) boundaries, while we use reflective conditions at the axis $r=0$. At the lower boundary $z=z_L=0.5$, for $r > \theta_j z_L$, we extend in the ghost cells the initial analytical pressure and density profiles, for ensuring dynamical equilibrium. We keep these values constant. The jet is injected in the region defined by $0.5 < z < 1$ and $0 < r < \theta_j z$, where we set the velocities, density and pressure defined in Eqs. \ref{eq:vz}, \ref{eq:vr}, \ref{eq:rho} and \ref{eq:prs}. In addition the fluxes of the Riemann solver are set to zero and hence the fluid variables remain unchanged in this region. This is done to avoid spurious effects at the lower boundary. 

The 2D case is evolved until a steady state is reached, up to $t_{f}=3000$ in units of $z_0/c$.

\subsection{3D setup}
The initial condition for the 3D simulation is the axisymmetric steady-state reached in  the 2D simulations described above. The 2D results, obtained in the cylindrical coordinates $(r,z)$, are projected on the Cartesian coordinates $(x,y,z)$.
The computational domain is made of $550\times550\times1850$ grid points, covering the physical domain $[-5,5]\times[-5,5]\times[1,30]$.  Like in the 2D case, the grid is uniform only in the inner region, $[-1,1]\times[-1,1],\times[1,20]$, with $300\times300\times1500$ cells, and stretched outside. 
We use outflow conditions at all boundaries, except at the lower one. In the 3D simulation we set the lower boundary  at $z=z_0=1$. For $\sqrt{x^2+y^2} < \theta_j \, z_0$ we have injection conditions with the jet parameters, and for $\sqrt{x^2+y^2} > \theta_j \, z_0$ we set density and pressure constant in time and following the initial profiles, as we have done in 2D. 
\subsection{Calculation of the average of the propagation velocity}
In order to average the velocity of the moving material, plotted in \ref{fig:entrainment}, we defined a step function $g(x,y,z)$ as
\begin{equation}
    g(x,y,z,t) = \begin{cases}
        0\quad\mbox{if }v_z(x,y,z,t)<0.1\\
        1\quad\mbox{if }v_z(x,y,z,t)\geq0.1
    \end{cases}
\end{equation}
that selects the jet section with a threshold on the propagation velocity $v_z$.
The average velocity $\langle v_z\rangle$ is then the average of the distribution of the velocity on the $x-y$ jet section, calculated as
\begin{equation}
    \langle v_z(z,t)\rangle= A^{-1}\int_{xy}v_z(x,y,z,t)g(x,y,z,t)dx dy\,,
\end{equation}
where $A=\int_{xy}g(x,y,z,t)dx dy$.
\end{document}